\documentclass[journal=nalefd,manuscript=communication]{achemso}

\usepackage[version=3]{mhchem} 

\author{Kathryn Todd}
\affiliation[Stanford University Department of Physics]
{Department of Physics, Stanford University, Stanford, California, 94305, USA}
\author{Hung-Tao Chou}
\affiliation[Department of Applied Physics]
{Department of Applied Physics, Stanford University, Stanford, California, 94305, USA}
\author{Sami Amasha}
\affiliation[Stanford University Department of Physics]
{Department of Physics, Stanford University, Stanford, California, 94305, USA}
\author{David Goldhaber-Gordon}
\email{goldhaber-gordon@stanford.edu}
\affiliation[Stanford University Department of Physics]
{Department of Physics, Stanford University, Stanford, California, 94305, USA}

\title[\texttt{Quantum Dots in Graphene Nanoconstrictions}]
{Quantum Dot Behavior in Graphene Nanoconstrictions}

\begin{document}
\begin{abstract}
Graphene nanoribbons display an imperfectly understood transport gap. We measure transport through nanoribbon devices of several lengths. In long ($\geq$250 nm) nanoribbons we observe transport through multiple quantum dots in series, while shorter ($\leq$60nm) constrictions display behavior characteristic of single and double quantum dots. New measurements indicate that dot size may scale with constriction width. We propose a model where transport occurs through quantum dots that are nucleated by background disorder potential in the presence of a confinement gap.
\end{abstract}

Graphene is a two-dimensional conductor with remarkable properties ranging from long spin relaxation times
\cite{Tombros2007} to very high mechanical strength \cite{Booth2008} to the
highest known room temperature mobilities \cite{Bolotin2008}. Extended graphene sheets also display a non-zero minimum conductivity even at nominally zero carrier densities, which limits their applicability for some types of semiconductor devices such as transistors with high on-off ratios.
When graphene is etched into narrow strips known as nanoribbons, however, conductance through the device is suppressed for a wide range of Fermi energies around the Dirac point. The observation of this conduction gap,\cite{Han2007, Chen2007, Li2008} which scales inversely with ribbon width,\cite{Han2007} has led to the proposal of graphene nanoribbons as next-generation high-frequency transistors. However, the observed conduction gaps are larger
than those expected from a single-particle confinement picture in the absence of lattice effects at the edges, while studies on nanoribbons of different orientations with respect to the graphene lattice\cite{Han2007} demonstrate that explanations depending on the presence of well-defined crystallographic edges\cite{Son2006, Brey2006, Yang2007} do not apply. Several alternative explanations have been put forward to
explain the large energy scale of the conduction gap. One proposal involves a non-conductive "dead zone" at
the ribbon edges due to atomic-scale disorder that causes the effective conducting ribbon width to be narrower
than the physical width\cite{Han2007}. Elaborations on this idea propose that atomic-scale disorder near the ribbon
edge may give rise to localized states that extend into the ribbon body\cite{Evaldsson2008}. Conversely, random charged
impurity centers in the body of the ribbon, in combination with the ribbon's confinement gap, have been proposed
to cause a metal-insulator transition\cite{Adam2008}. Others have suggested that lithographic line edge roughness may
create a series of quantum dots defined by nanometer-scale variations in ribbon width\cite{Sols2007}. Here we
examine detailed conduction characteristics of long nanoribbons as well as short nanoconstrictions in an attempt
to elucidate the origin of the large conduction gap in graphene nanoribbons. In light of our data, we propose that
the random charged impurity centers present throughout the graphene sheet, in conjunction with a gap stemming from the constriction's confined
geometry, give rise to isolated puddles of charge carriers acting as quantum dots both in long nanoribbons and
in short nanoconstrictions, even in cases where the constriction length is so short that there is no hard transport gap. The physical origin of quantum dot behavior in graphene nanoribbons has implications for the manufacture of graphene nanoribbon transistors.

\begin{figure}
  \includegraphics{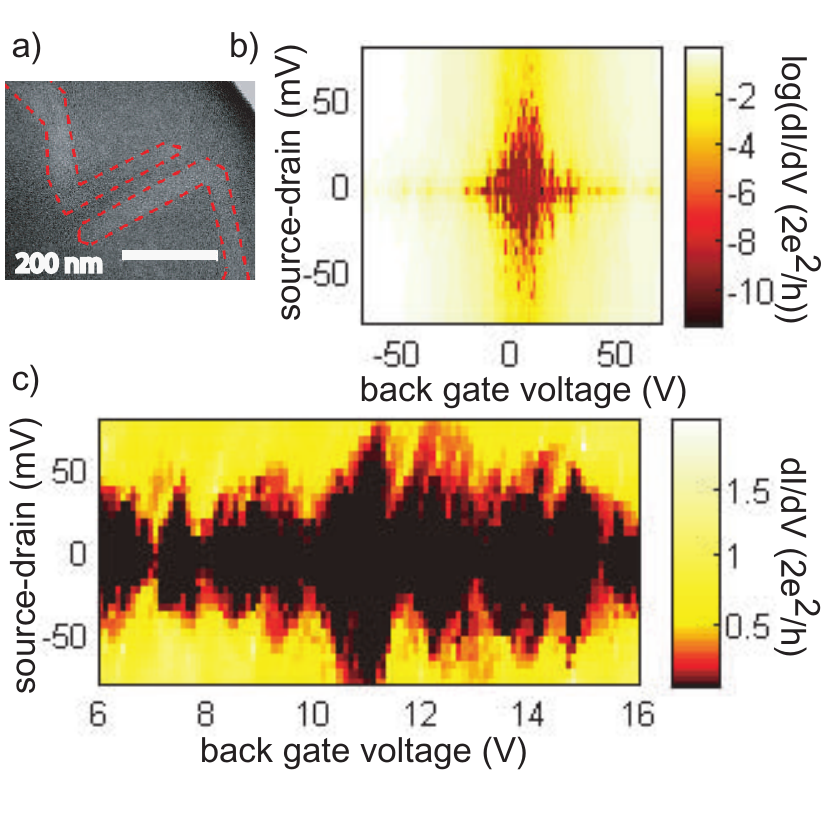}
  \caption{a) SEM micrograph of a long graphene nanoribbon 250 nm in length and 40 nm in width. Lighter areas are where graphene has been removed by an oxygen plasma etch to define the constriction. Dashed red lines are guides to the eye indicating the boundary of the etched area. b) Logarithmic map of the differential conductance dI/dV versus back gate voltage and source-drain bias acquired at 4.2 K showing conduction gap. c) dI/dV over a smaller range, showing Coulomb blockade behavior suggesting the presence of several quantum dots in series.}
  \label{fig:fig1}
\end{figure}

As reported elsewhere \cite{Han2007,Chen2007}, long ($\geq$1 micron) graphene nanoconstrictions display gapped behavior: conduction
is suppressed by several orders of magnitude for a wide range of gate voltages around the Dirac point,  and for tens
of mV of source-drain bias. In \ref{fig:fig1} we show data from a 250 nm long, 40 nm wide graphene nanoribbon patterned
by oxygen plasma etching through a PMMA mask on a graphene flake. This flake and all others discussed in this paper have been verified to be single-layer by Raman spectroscopy\cite{Ferrari2006,Graf2007}, and are contacted by Ti/Au metal patterned far from the constriction region. We estimate the flake mobility to be 500 cm$^{\textrm{2}}$/Vs based on two-wire measurements made between contacts not spanning the constriction.\bibnote{Because two-wire measurements do not allow us to eliminate the effects of contact resistance, we base our mobility estimates on the slope of conductance traces at relatively low densities 10 V from the Dirac point where the graphene sheet resistance is still dominant over the contact resistances.} All data reported herein are from two-probe measurements: in general the constriction resistance is larger than any other resistance in the circuit, and in any case four-probe measurements cannot eliminate the resistance of the region of the graphene flake between the metal contacts and the constriction region. We see the large scale gap reported by others (Figure 1b).\cite{Han2007,Chen2007} Careful examination of the
conduction inside the gapped region (Figure 1c) reveals Coulomb diamond-like features, with conduction suppressed by more than 7 orders of magnitude around zero source-drain voltage
for nearly all gate voltages, indicating the presence of several quantum dots in series. From the dI/dV map we calculate
dot areas ranging from 860 to 1700 nm$^{\textrm{2}}$, consistent with the formation of dots that span the width of the ribbon, with lengths of the same order as their widths. \bibnote{We calculate dot size based on a parallel plate capacitor model where the back
gate defines one plate of the capacitor and the graphene sheet and nearby metal gates define the other plate, justified by
the presence of conducting graphene sheet everywhere except for the two narrow strips defining the constriction. The presence
in each of the samples discussed here of graphene or metal gates near the constriction may result in some screening of the
back gate voltage, resulting in an underestimation of dot areas in each case. In data sets where dI/dV is measured with
respect to side gate voltage, we calculate an equivalent back gate voltage using a separately measured multiplier.} In the course of the paper we use the same method to calculate dot sizes in nanoconstrictions of varying width and length. We summarize the sample dimensions, mobilities and dot sizes in \ref{tbl:sizes}.

\begin{figure}
  \includegraphics{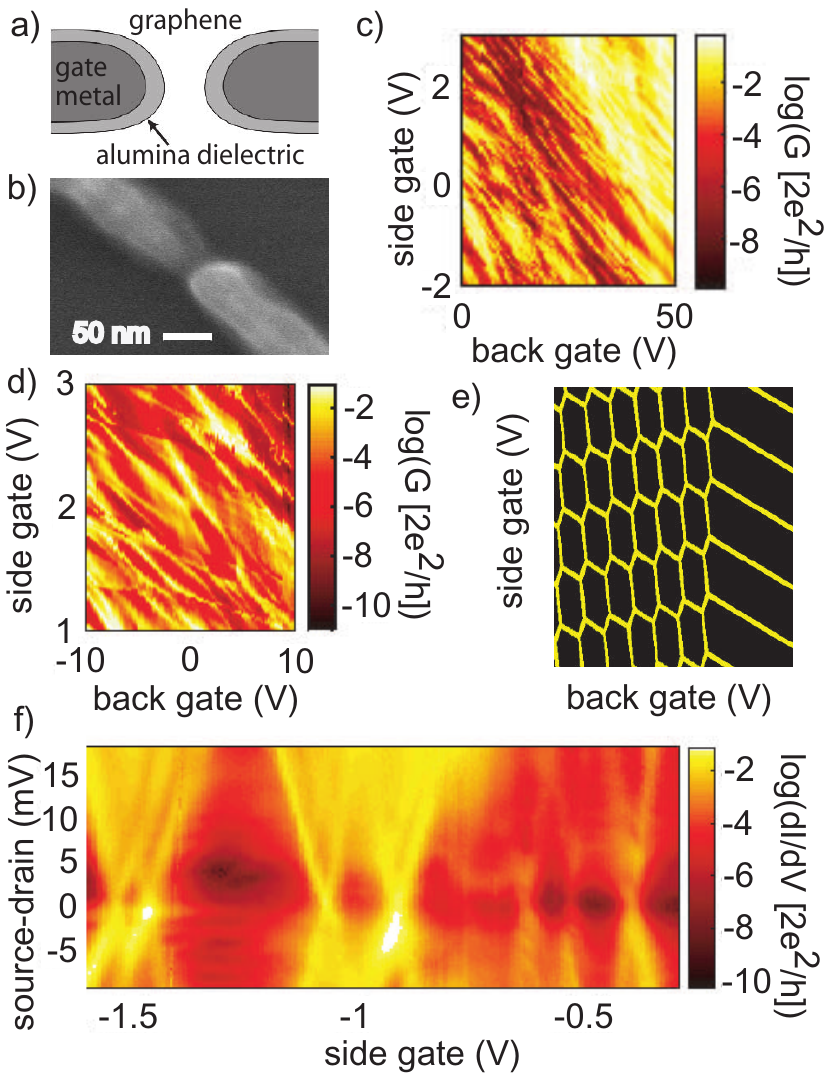}
  \caption{a) Schematic of the device geometry of our short nanoconstrictions. b) SEM micrograph of the 60 nm long, 15 nm wide constriction, showing the gate metal (bright regions), the alumina dielectric (transparent bright regions) and the graphene flake (dark regions). c) Stability diagram showing the logarithm of the conductance versus back gate and left side gate voltages. This and all data sets described in this figure were taken at 4.2 K. d) A higher resolution stability diagram showing the transition from single to double-quantum dot behavior. A few switching events, likely caused by trapped charges moving in the dielectrics cause translations in the pattern. e) Cartoon illustrating charge transitions in an ideal double-dot system versus the voltages of two gates that couple asymmetrically to each dot. In a parallel-coupled double-dot system with strong coupling to the leads, yellow lines would correspond to conductance maxima. A transition to single-dot behavior in the right side of the panel occurs as one dot is fully depleted. f) Logarithmic dI/dV versus left side gate voltage, for a constant back gate voltage of 30 V. At this back gate voltage we expect the Dirac point to lie at V$_{\textrm{sg}}$ $\sim$ 0.}
  \label{fig:fig2}
\end{figure}

In order to better understand the origin of these quantum dots in long nanoribbons, we have fabricated a series of shorter ($\leq$60 nm) constrictions: in doing so we hoped to isolate a small number of dots in the constriction region and study their behavior in detail. Fabrication of our short constrictions begins as for our long constrictions with the deposition of Ti/Au contacts on a graphene flake far from the constriction region and an oxygen plasma etch through a PMMA mask to define the constriction. In most of our short constrictions, however, this step is followed by the deposition of an aluminum oxide insulator and then gate metal, and finally liftoff, resulting in a nanoconstriction situated between two
separately addressable side gates. In \ref{fig:fig2} we show data from a 60 nm long, 15 nm wide constriction. We estimate the mobility of the flake far from the constriction region to be 700 cm$^{\textrm{2}}$/Vs, although we emphasize that the processing steps that the constriction region has undergone in our side-gated constrictions may cause its mobility to differ from the mobility of the flake as a whole. Conductance traces with respect to back gate voltage show a wide region of suppressed
conductance punctuated by narrow conductance peaks. A stability diagram (Figure 2c) of conductance versus back gate and
one side gate exhibits features moving with a constant slope. From this slope we extract a measure of the relative capacitance of the side and back gates to the graphene in the constriction region. We identify the diagonal region of lowest
average conductance running through the center of figure 2c as the set of back and side gate voltage pairs that bring the
potential inside the constriction to the Dirac point. Within this diagonal region we observe relatively narrow and straight
conductance peaks running in parallel. This region of simple conductance peaks, where conductance depends only on a single variable (the Fermi energy inside the constriction), is indicative of the presence of a single quantum dot, while outside the central Dirac
region (see zoom in, Figure 2d) conductance peaks bend and exhibit hexagonal patterns characteristic of two or more quantum
dots. We include a cartoon of the pattern expected from an idealized double-quantum dot system with a transition to single-dot behavior (Figure 2e).\bibnote{The model we use to generate the expected double-dot behavior does not take into account the change in Fermi level with gate voltage that takes place in our system.}

We measure nonlinear conductance across the constriction versus side gate voltage and source-drain bias (Figure 2f), setting the back gate to a constant
30 V so that the side gate accesses the Dirac point at approximately 0 V. At side gate voltages V$_{\textrm{sg}}$ near 0 V, we observe
quasi-periodic Coulomb diamonds which all share similar heights in source-drain voltage, indicating that each diamond
is produced by a state with the same total capacitance: in other words, reflecting the presence of a single quantum dot.
At values of V$_{\textrm{sg}}$ below -1.1 V, a side gate voltage similar to that at which we observe a transition from single- to double-dot behavior in
Figure 2d, the differential conductance pattern changes: we observe tall diamonds that intersect to form a small diamond at V$_{\textrm{sg}}$ = -1.5 V, suggesting that
conduction is occurring through two quantum dots in parallel. From the spacing in gate
voltage of the conductance peaks defining the uniform diamonds near the Dirac point and from the larger diamonds at higher V$_{\textrm{sg}}$ we calculate dot areas of
900 nm$^{\textrm{2}}$ +/-70 nm$^{\textrm{2}}$ for the single, larger dot near the Dirac point and 300 nm$^{\textrm{2}}$ +/- 20 nm$^{\textrm{2}}$ for the second, smaller dot
at larger gate voltages \bibnote{The uncertainties come largely from the conversion from side gate to equivalent back
gate voltage}. Excited state features with energy spacing of ~7 meV can be seen running in parallel with the diamond
boundaries in the diamond centered at V$_{\textrm{sg}}$  =  -1.0 V in Figure 2f, along with inelastic cotunneling features in the diamond centered at -1.3 V
(horizontal lines spanning the diamond). Excited state features as well as inelastic cotunneling features demonstrate the presence of two distinct energy scales. The existence of two distinct energy scales in our diamonds is enough alone to establish that these features are not conduction resonances stemming from transport through a single tunnel barrier. To rule out the possibility that the features are due to Fabry-Perot-like transport through double- or multiple barriers, we consider the energy scales involved. In a system such as this, with very low average conductance and therefore opaque tunnel barriers, Fabry-Perot transport would be equivalent to transport through a very large confined area, one so large that the charging energy has become negligible. The energy scales we observe in our device, however, are non-negligible: our diamond heights ($\sim$ 4-20 mV) are in fact larger than those measured in many lithographically defined graphene quantum dots \cite{Ponomarenko2008, Stampfer2008} where the existence of a non-negligible charging energy is well-established. We therefore conclude that the two energy scales we observe are the charging energy required to add an additional electron to the dot and the spacing of confinement energy levels inside the dot. The presence of inelastic cotunneling features is a further indication of transport through a quantum dot, as they stem from energy dissipation occurring inside the feature responsible for the conduction peaks, which is not possible inside a single simple tunnel barrier.

The dot size derived from our measured level spacing is inconsistent with energy level transitions in a circular quantum dot of the area calculated from the measured charging energy. Unlike experiments where quantum dots in graphene are geometrically defined\cite{Stampfer2008, Ponomarenko2008}, we
have no measure of the shape (as opposed to the area) of the dot under study, and no reason to assume that it has a
well-defined circular or square geometry. For more realistic geometries the Dirac spectrum produces chaotic level spacings from which
it is not possible to extract geometric information without detailed statistics \cite{deRaedt2008, Ponomarenko2008}. From the side gate voltage where the larger diamond first appears, though, we estimate that the lowest energy level of this dot is located 100 meV from the Dirac point. We note that
the bottom of the potential defining the dot does not necessarily coincide with the spatially averaged Dirac point measured
by transport: however, should it be the case that the bottom of the potential defining the dot is not far from the spatially averaged Dirac point we calculate that this energy level corresponds to a length scale of 21 nm, which is in rough agreement with the size of this smaller quantum dot calculated from the spacings of Coulomb blockade peaks.

In order to gain information about localization lengths in narrow constrictions, we have fabricated on the same
graphene flake a constriction of similar (10 nm) width but of 30 nm length. This very short constriction fails to show
highly suppressed conductance near the Dirac point (see Figure 1 of the supplementary information). This is similar to the findings of Ponomarenko \emph{et al.} \cite{Ponomarenko2008} regarding the necessity of fabricating very short constrictions at the entrances of their lithographically patterned graphene quantum dots in order to ensure large coupling to the leads. Based on our findings, we estimate that the localization
length in our very narrow constrictions is on the order of 50 nm.

\begin{figure}
  \includegraphics{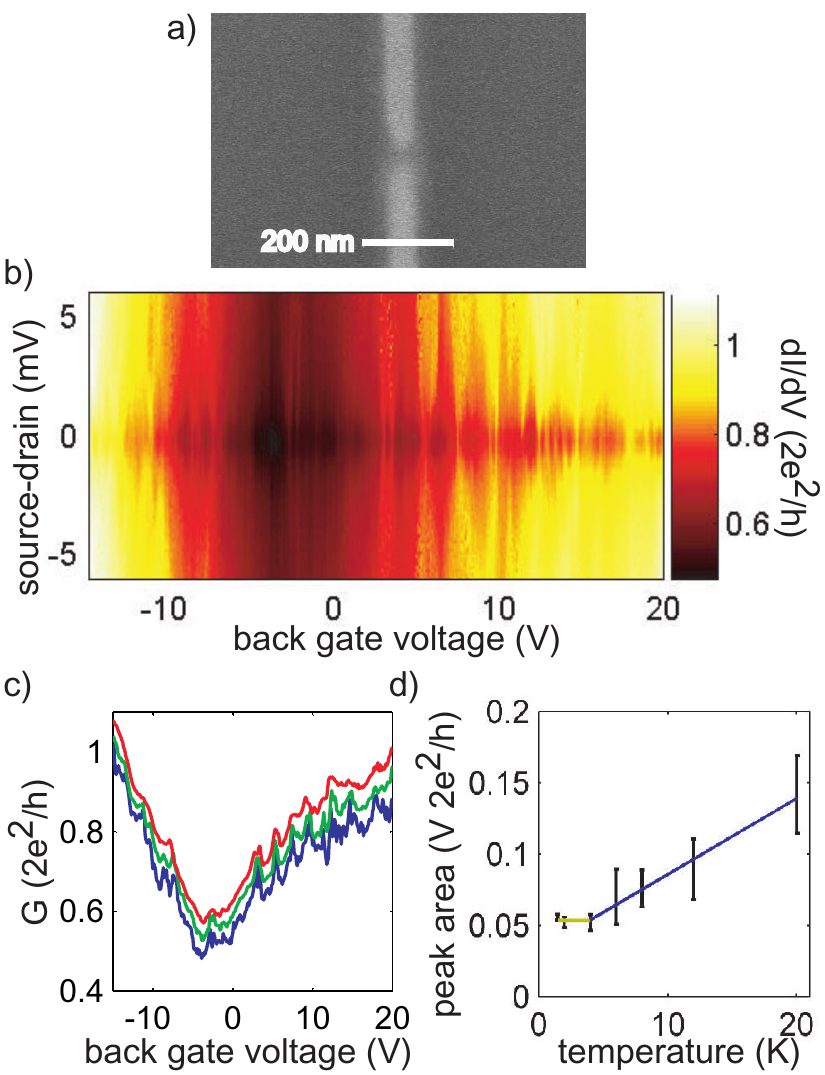}
  \caption{a) SEM micrograph of our 60 nm long, 35 nm wide constriction. b) Nonlinear conductance versus back gate voltage and source-drain bias for this device acquired at 1.5 K. c) Conductance through the constriction at (from bottom) 1.5 K, 6 K, and 20 K showing persistence of conductance peaks and d) Temperature dependence of the area under conductance peaks, showing a transition between constant peak area and a linear increase. Peak areas are calculated from fits to conductance peaks located at 6 V$_{\textrm{bg}}$, and one-sigma error bars are calculated from the uncertainties associated with those fits. Colored lines are guides to the eye, indicating the transition between constant peak area in the low temperature single-level transport regime and linearly increasing peak area in the high temperature multi-level transport regime.
}
  \label{fig:fig3}
\end{figure}

In contrast to the narrowest short constrictions, in short 60 nm long constrictions with widths of 35 and 55 nm, a
regime where long nanoribbons (in work by us and others) \cite{Han2007, Chen2007} show gapped conduction behavior at low
temperatures, we do not observe a hard gap in transport. Instead, we measure conduction profiles similar to those observed in extended graphene sheets but with Coulomb diamond features overlaid. \ref{fig:fig3} shows data obtained from the 60 nm long, 35 nm wide constriction, patterned on a flake whose mobility we estimate to be 3000 cm$^{\textrm{2}}$/Vs. We address the question of whether the diamond
features we observe reflect the presence of quantum dots or simply resonant tunneling through a barrier. In figure 3d we show
the temperature dependence of the conductance peak area for the peak occurring at ~6 V back gate voltage (V$_{\textrm{bg}}$). In transport through a quantum dot
a transition from a peak area that is constant in temperature to a peak area that increases linearly with increasing temperature
should be observed when kT becomes comparable to the level spacing in the dot, corresponding to the transition between transport through single and multiple energy levels in the dot. Such
a transition would also be observed in the case of resonant tunneling through a barrier, but it would occur at a temperature
equivalent to the level spacing within the barrier: there would be no difference between the temperature at which this
transition occurred and the "charging" energy measured in dI/dV. In our data, the temperature at which the transition from
single-level to multi-level transport occurs is on the order of 4 K, or ~0.7 meV, while the charging energy is greater than 5 meV for the peak in question. Another possibility is that the features we see in our source-drain maps reflect not Coulomb blockade but
Fabry-Perot behavior. In fact we do observe Fabry-Perot features similar to those observed by Miao et al \cite{Miao2007} but these features are much narrower than the Coulomb blockade features under discussion, disappear at
temperatures above 1.5 K, and are also present in measurements made across contacts not spanning the constriction (see Figure 2 of the supplementary information). Because of these factors, we conclude that these small modulations on our large-scale Coulomb blockade features are associated with Fabry-Perot resonances occurring in the unpatterned graphene sheet between our contacts and the constriction. Having eliminated single-barrier and Fabry-Perot resonances as likely causes of the conductance patterns we observe in this constriction, we conclude that we are measuring conduction through a quantum dot.

 Calculation of the dot areas from conductance peak spacing in the 60 nm long, 35 nm wide constriction gives areas ranging from 400-600 nm$^{\textrm{2}}$,
 while the regular periodicity of the diamonds suggests single dot behavior in this region near the Dirac point.
 Diamond heights increase as the back gate voltage approaches the Dirac point, reflecting either a change in the
 capacitance of the dot to the source and drain due to the dot growing physically smaller as the Fermi energy
 approaches the Dirac point, or due to a change in the quantum capacitance of the dot as the density of states
 decreases near the Dirac point. Similar behavior is seen in a 60 nm long, 55 nm wide constriction (see Figure 3 of the supplementary
 information), where we find dot areas ranging from 1600 to 7100 nm$^{\textrm{2}}$.  In \ref{tbl:sizes} we summarize the calculated dot areas
 for all constrictions discussed here. Although more measurements would be necessary to make strong quantitative statements, the available data suggest that the area of quantum dots scales with the width but not the length
 of the constriction. We do not rule out the possibility that sample mobility may also play a role in determining dot size.

We emphasize that the 60 nm long, 35 nm wide constriction is narrower than the 250 nm long, 40 nm wide nanoribbon described
in Figure 1, and yet displays Coulomb diamonds on top of a background that increases with gate voltage in a manner similar
to bulk graphene instead of the hard-gapped conduction seen in long nanoribbons \cite{Han2007, Chen2007}. This increasing
background provides a possible explanation for why we do not observe a transition from single to multiple quantum dot behavior
in wider constrictions: in these constrictions Coulomb blockade features at larger gate voltages would be obscured by the
rising Dirac background.

The short length of our constrictions and their geometry (where the constriction is narrowest at its center) makes it implausible that quantum dot behavior in these constrictions stems
from lithographic line edge roughness where the width of the ribbon varies randomly by a few nanometers over a length scale of tens of nanometers as has been proposed for long nanoribbons\cite{Sols2007}.  In addition, we observe changes in differential conductance data (see Figure 4 of the supplementary information) after thermal
cycling: the pattern of Coulomb diamonds becomes more complicated, suggesting multi- rather than single-dot transport, and
the inferred dot size changes by factors of $\sim$ 2. These changes imply that a model where Coulomb blockade behavior is produced
by line edge roughness alone does not apply.

\begin{figure}
  \includegraphics{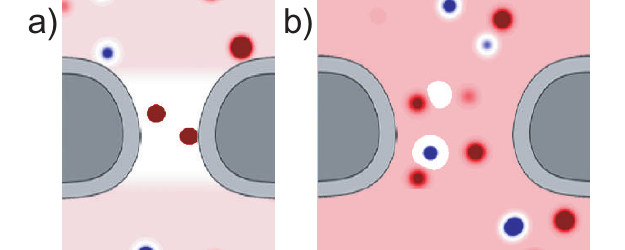}
  \caption{Cartoon rendering of the possible effect of background potential fluctuations on charge carrier density near the Dirac point in a graphene nanoconstriction device. Outside the constriction islands of more heavily p- (blue) or n-doped (red) regions are embedded in a uniform charge density determined by the gate voltage. Inside the constriction, the presence of a confinement gap results in regions wholly empty of charge carriers (white) surrounding small islands of charge carriers at the most heavily doped points. a) At gate voltages near the Dirac point and narrow constriction widths the constriction region is empty of charge carriers except at the locations of strongest local potential b) At higher gate voltages and/or wider constriction widths charge carriers can move freely across the constriction, which contains a small number of isolated dots.
}
  \label{fig:fig4}
\end{figure}

Instead we propose that the quantum dot behavior we observe in our samples may stem from the doping inhomogeneity of bulk graphene on \ce{SiO2}\cite{Martin2008}, combined with the presence of a confinement gap from the constriction. The combination of these two effects would generate carrier-free regions at gate voltages near the Dirac point punctuated by islands of p- and n-type carriers: in \ref{fig:fig4} we show a cartoon based on simulations of the electron density induced by a gate voltage in combination with a random impurity potential landscape, where impurity density is extracted from sample mobility\cite{Chen2008, Ando2006, Nomura2007, Cheianov2007, Hwang2007}. Unlike more sophisticated treatments\cite{Rossi2008, Lewenkopf2008} our cartoon model does not take into account electron interaction effects in calculating the local electron density. After calculating the local potential due to the charged impurity landscape we impose a gap based on the confinement energy due the constriction and set the electron density to zero whenever the Fermi level plus the local potential is less than the gap. This procedure results in isolated puddles of electrons and holes embedded in an insulating environment for small gate voltages and narrow constrictions (Figure 4a) and a combination of isolated puddles and conducting regions spanning the constriction length for gate voltages farther from the Dirac point or wider constriction widths (Figure 4b). For more information about the generation of these cartoons, see the supplementary information. Experimental indications that dot size roughly scales with constriction width, but not with constriction length are consistent with our proposed model, and are inconsistent with the edge-defect model in a regime where localized states do not span the entire constriction width. It is likely that bulk mobility also plays a role in the size of dots formed in this scenario: future experiments will provide more quantitative information on the relationship between dot size, mobility and constriction width. In contrast to our model where bulk disorder is the most important factor in determining conductance properties of graphene nanoconstrictions, conduction gaps in long nanoribbons have in the past been explained in terms of Anderson localization caused by edge defects
\cite{Evaldsson2008, Mucciolo2008}.  In our short constrictions, the presence of nearby metal gates seems to pin the Dirac point in the
ribbon at zero gate voltage, in contrast to our devices free of patterned metal gates where samples are generally p-type at zero gate voltage.
Pinning, or alternatively screening effects from the nearby metal, may decrease the importance of edge disorder in our side-gated
samples. In support of this proposal is the fact that data from samples fabricated in the same short-constriction geometry as those described above
but without the deposition of metal gates on top of the etched area shows disordered differential conductance through the constriction with few identifiable Coulomb
diamonds (see Figure 5 of supplementary information for an example of such a sample). We also emphasize that our measurements on short constrictions fabricated without metal side gates imply that our model, where bulk disorder effects dominate over edge effects, may be most applicable in devices with nearby patterned metal side gates. While we believe that bulk disorder should affect conductance through all constrictions, in constrictions fabricated without nearby metal gates effects associated with disordered edges might be more important. Local probe studies such as scanning gate microscopy on short constrictions patterned without metal side gates may in the future provide more information on the relative importance of bulk and edge disorder in non-side-gated constrictions and nanoribbons.

\suppinfo
Supporting information is available: we describe our methods for generating the cartoons depicted in Figure 4, and we provide figures describing transport behavior through a 30 nm long, 10 nm wide constriction, low temperature Fabry-Perot behavior of a 60 nm long, 35 nm wide constriction, transport behavior of a 60 nm long, 35 nm wide constriction after thermal cycling, and transport behavior of a short constriction fabricated without the deposition of metal gates in the etched area defining the constriction. This material is available free of charge via the Internet at http://pubs.acs.org.

\acknowledgement

The authors thank Mark Brongersma for access to a Raman spectroscopy system, Joey Sulpizio for help with Raman spectroscopy, and Xinglan Liu, Shaffique Adam, and Eduardo Mucciolo for helpful discussions. This work was supported in part by AFOSR grant FA9550-08-1-0427 and by the MARCO/FENA program. K.T. acknowledges support from the Intel and Hertz Foundations. H.-T. C. acknowledges support from the David and Lucile Packard Foundation. Work was performed in part at the Stanford Nanofabrication Facility of NNIN supported by the National Science Foundation under Grant
ECS-9731293.

During the preparation of this manuscript we became aware of work on multiply-gated long nanoribbons that reaches similar conclusions about the source of quantum dot behavior in graphene nanoribbons.\cite{Stampfer20082}

\begin{table}
  \caption{Dot characteristics and constriction width: constriction width and length are determined from SEM micrographs taken of the device after measurement. Dot areas are calculated from Coulomb diamond geometry. Variation comes from both from errorbars in the calculation and also from the presence of diamonds of different widths. Equivalent diameters are extracted from calculated dot areas assuming circular geometry, and are presented only to allow a rough comparison with constriction width. Flake mobilities are estimated from two-wire measurements made between contacts on a single side of the constriction, and impurity densities are estimated from flake mobilities.}
  \label{tbl:sizes}
  \begin{tabular}{llllll}
    \hline
     width & length & dot areas & equivalent diameter & flake mobility & impurity density \\
    \hline
    15 nm & 60 nm & 300 to 900 nm$^{\textrm{2}}$ & 20 to 34 nm & 700 cm$^{\textrm{2}}$/Vs & 7 x 10$^{\textrm{12}}$ cm$^{\textrm{-2}}$\\
    \hline
    35 nm & 60 nm & 400 to 600 nm$^{\textrm{2}}$ & 23 to 28 nm & 3000 cm$^{\textrm{2}}$/Vs & 2 x 10$^{\textrm{12}}$ cm$^{\textrm{-2}}$\\
    \hline
    40 nm & 250 nm & 860 to 1700 nm$^{\textrm{2}}$ & 33 to 47 nm & 500 cm$^{\textrm{2}}$/Vs & 10 x 10$^{\textrm{12}}$ cm$^{\textrm{-2}}$\\
    \hline
    55 nm & 60 nm & 1600 to 7100 nm$^{\textrm{2}}$ & 45 to 95 nm & 700 cm$^{\textrm{2}}$/Vs & 7 x 10$^{\textrm{12}}$ cm$^{\textrm{-2}}$\\
    \hline
  \end{tabular}
\end{table}

\newpage
\section{Supplementary Information}
\subsection{Simulations}
To generate the cartoons featured in figure 4 of the main paper, we generate a set of random impurities at density per lattice site\cite{Ando2006, Nomura2007, Cheianov2007, Hwang2007} \[n_{imp} = \frac{C}{\mu a}\] with strengths distributed uniformly over the energy interval $[-\delta, \delta]$\cite{Rycerz2007}, where we choose \cite{Chen2008} $C = 5 x 10^{15}$ and \[ \delta = t (\frac{a}{\xi})^{2} \sqrt{\frac{K_{0}}{40.5 n_{imp} a^{2}}} \] where t is the nearest-neighbor hopping energy $\approx 2.7 eV$, $\xi$ is the screening length in the material, which we choose to be $4 a$ following Lewenkopf \cite{Lewenkopf2008} and \[K_{0} = \frac{2 \lambda_{F}}{\pi \lambda_{mfp}}\] We calculate the local potential at every point on our mesh due to the presence of all of the charged impurities, and then employ a crude method that neglects electron interaction effects to get a rough measure of the local density at each point $r$ due to the charged impurities and Fermi energy due to the overall back gate voltage: \[n_{e}(r) = sign(E_{F} + r_{s} V(r))\left(\frac{E_{F} + V(r) r_{s}}{\hbar v_{F}}\right)^{2}\] where $r_{s}$ the coupling constant on the \ce{SiO2} substrate\cite{Rossi2008} $= 0.8$
Finally, we set the density to zero whenever \[\left|E_{F} + V(r) r_{s}\right| \leq E_{gap} \] where $E_{gap} = \hbar v_{F} \frac{\pi}{w}$ and $w$ is the width of the constriction. This results in constrictions completely empty of charge carriers except at the locations of largest $V(r)$ for low Fermi energies or narrow constrictions, and constrictions where small regions of charge carriers are isolated from a conducting sea by small annuli empty of charge carriers for higher Fermi energies and wider constrictions.
\newpage

\begin{figure}
  \includegraphics{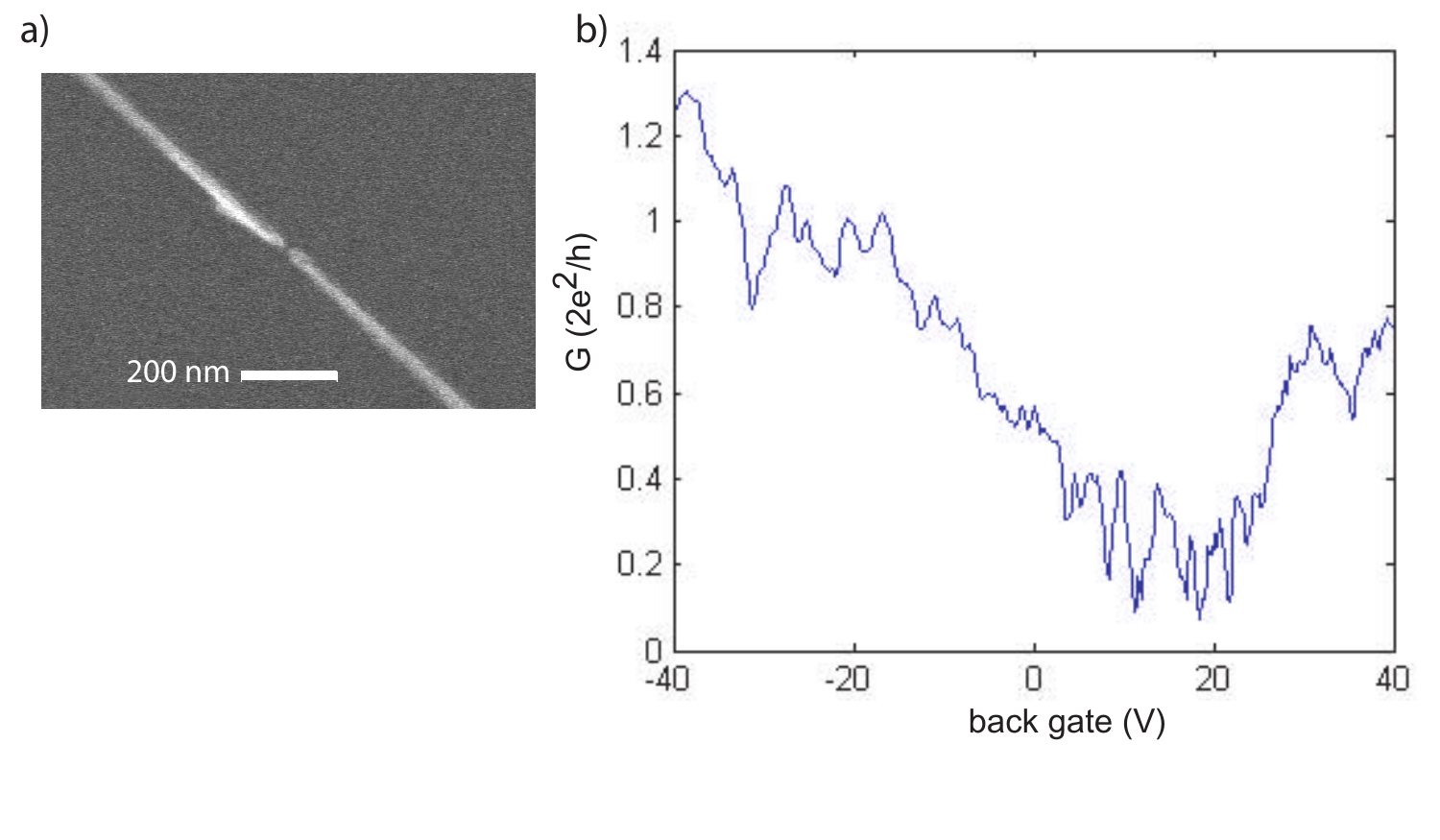}
  \caption{a) SEM micrograph of the 30 nm long, 10 nm wide constriction fabricated on the same flake as the constriction discussed in Figure 2 of the main paper. Despite the fact that it is very narrow, this short constriction displays b) high overall conduction and shows no gap around the Dirac point. Data acquired at 4.2 K.}
  \label{fig:suppS1}
\end{figure}

\begin{figure}
  \includegraphics{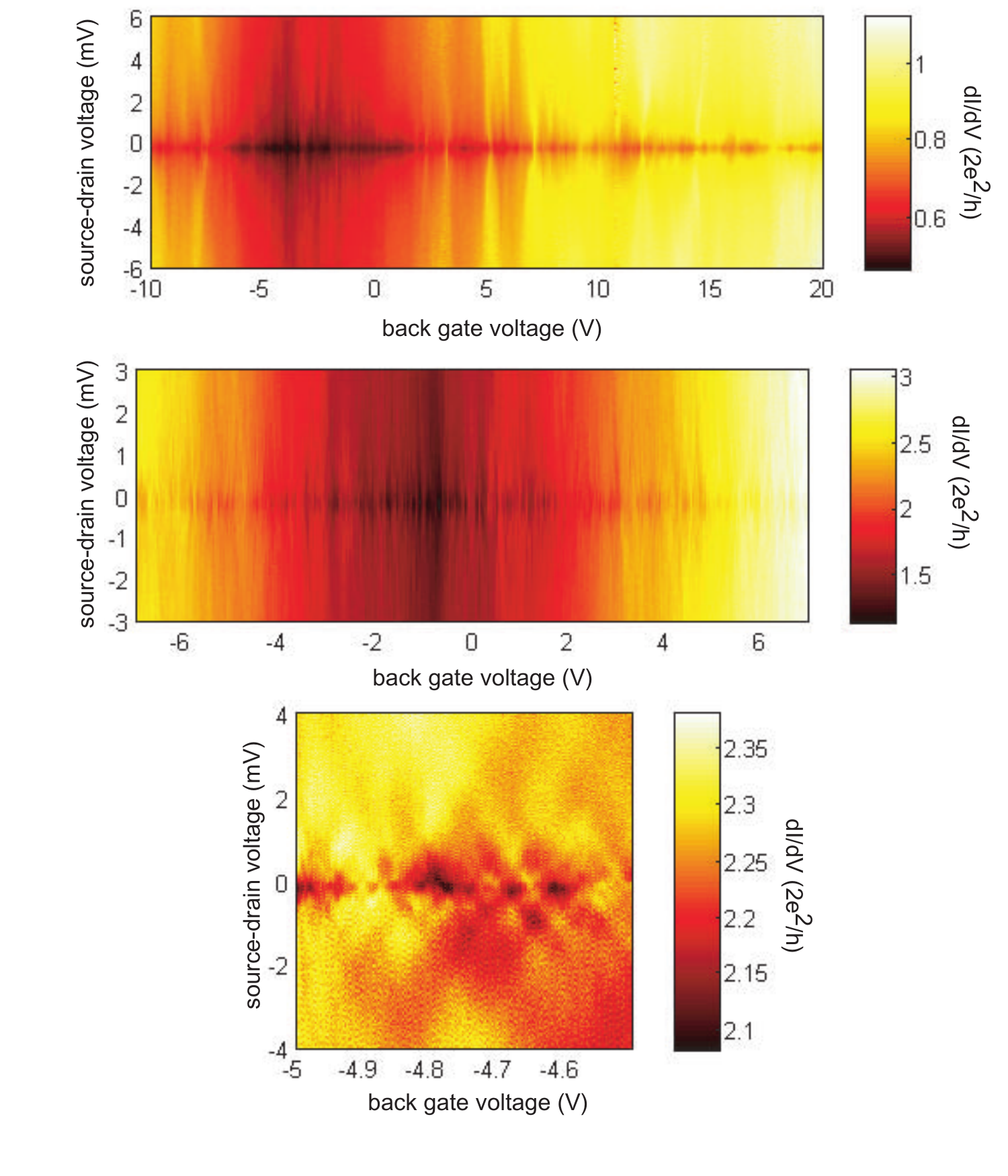}
  \caption{a) Nonlinear conductance map from the 60 nm long, 35 nm wide constriction described in figure 3 of main paper taken at 250 mK. At this low temperature narrow features are overlaid on the wider Coulomb diamond features seen also at higher temperatures (see Figure 3 of the main paper) b) Nonlinear conductance map taken at 250 mK across two contacts located on the same side of the constriction on the same sample. Narrow features are also present in this data set, showing that these features are independent of the presence of a constriction c) High resolution data set of the same features seen in b). At high resolution these features resemble Fabry-Perot resonances between sample contacts separated by micron length scales}
  \label{fig:suppS2}
\end{figure}

\begin{figure}
  \includegraphics{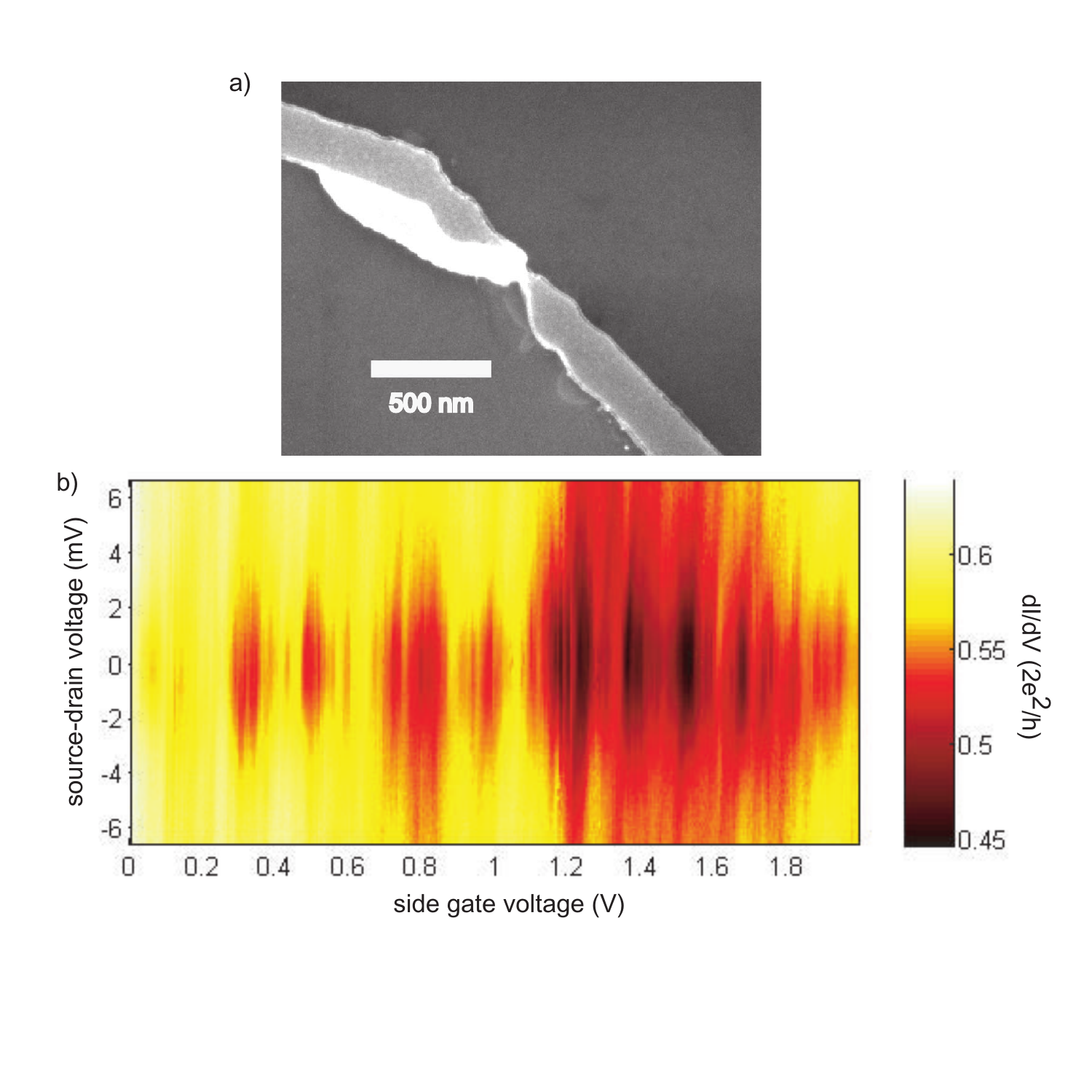}
  \caption{a) SEM micrograph of a 60 nm long, 55 nm wide constriction. Bright white material between the side gates is aluminum oxide that failed to lift off during fabrication. A measurement of conductance between the two side gates confirms that there is no metal shorting the constriction b) Nonlinear conductance across the constriction versus side gate voltage and source-drain bias taken at 4.2 K. Coulomb diamonds are visible on top of a large background conductance, as in the 60 nm long, 35 nm wide constriction described in Figure 3 of the main paper
}
  \label{fig:suppS3}
\end{figure}

\begin{figure}
  \includegraphics{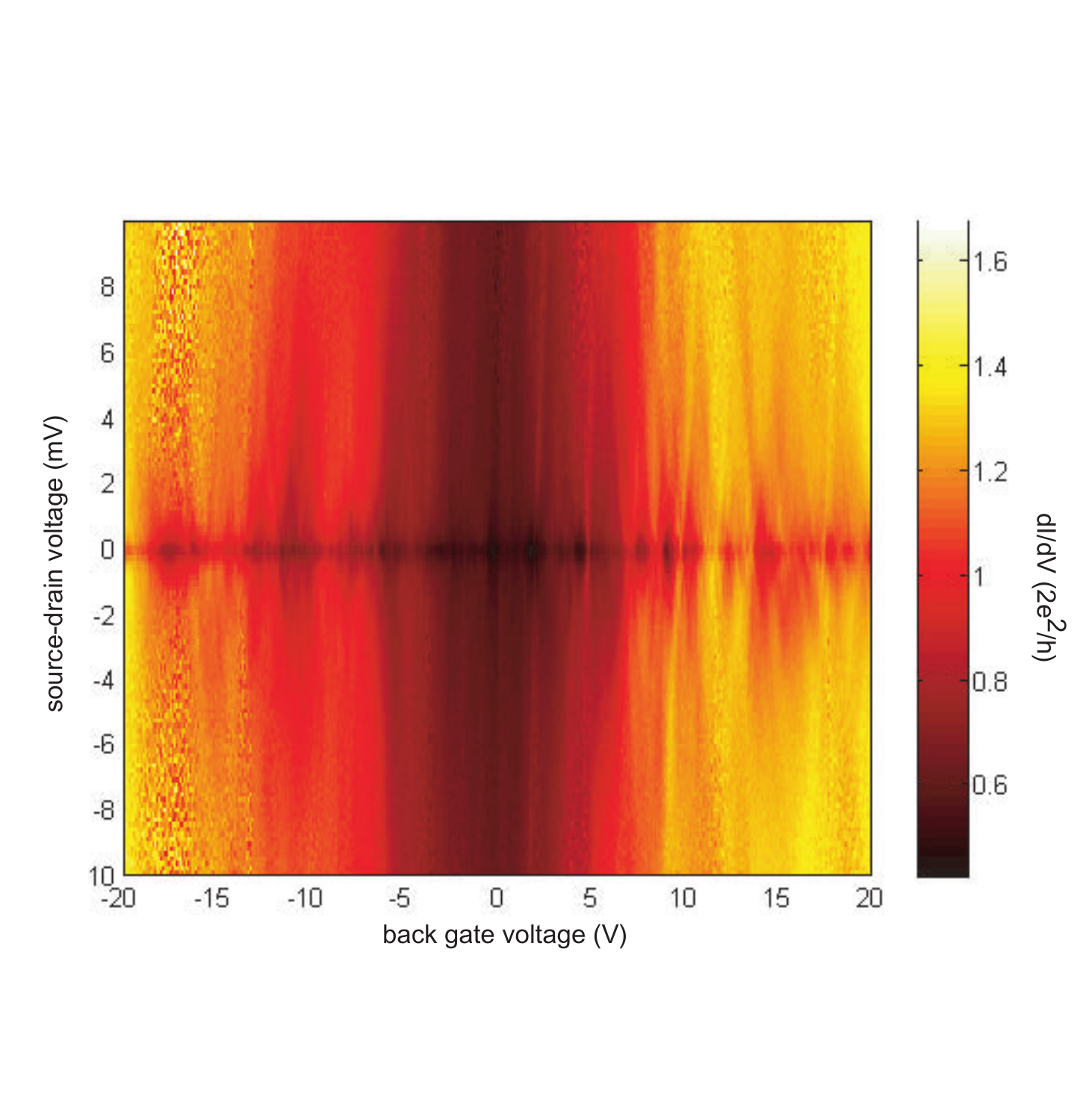}
  \caption{a) Nonlinear conductance map of 60 nm long, 35 nm wide constriction described in figure 3 of main paper at 250 mK after thermal cycling. The pattern of Coulomb diamonds has become less regular, and dot areas calculated from diamond widths have changed by factors as large as 1.75.
}
  \label{fig:suppS4}
\end{figure}

\begin{figure}
  \includegraphics{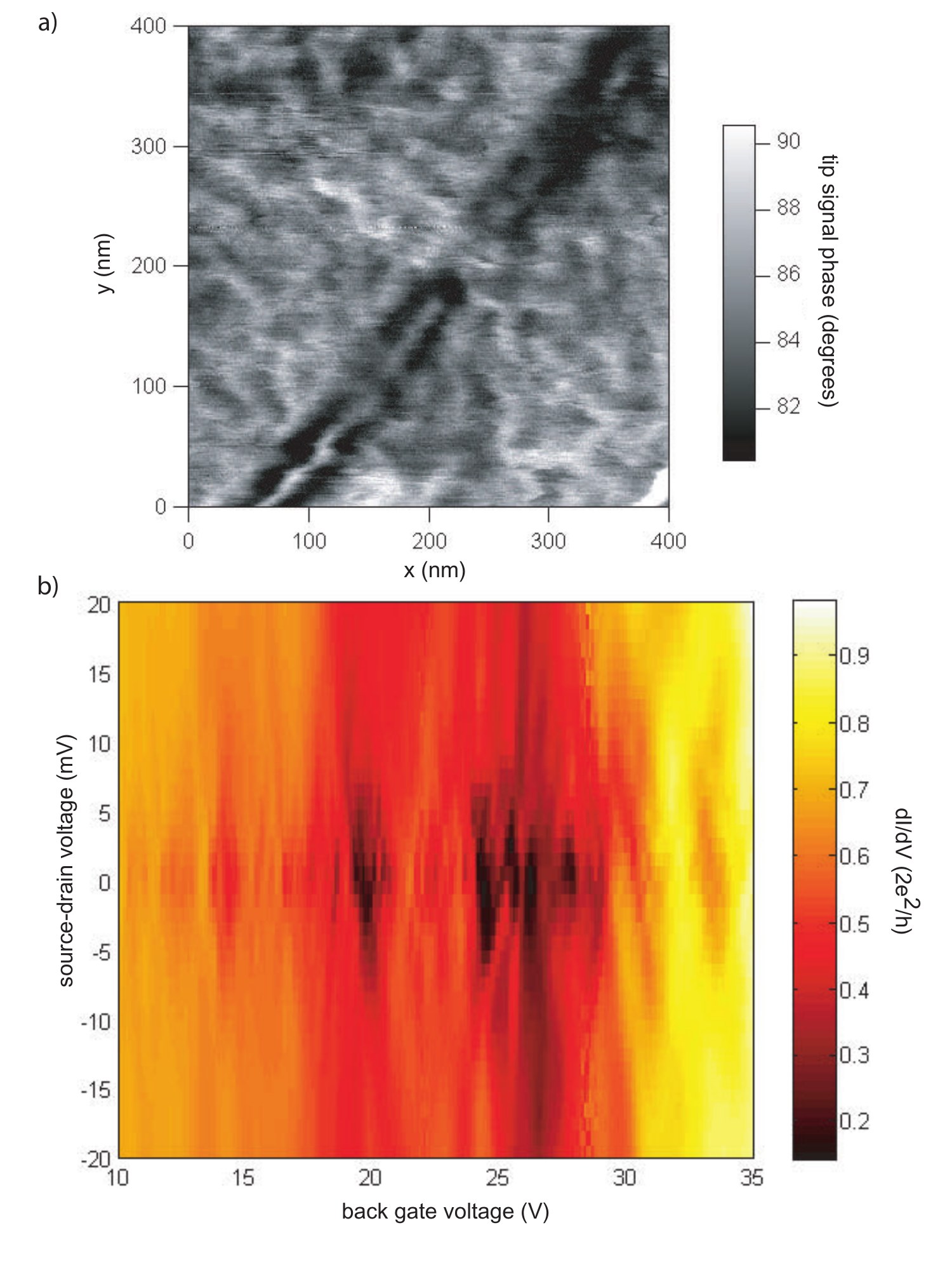}
  \caption{a) AFM micrograph of 35 nm long by 40 nm wide constriction fabricated without the deposition of metal gates on top of the etched area defining the constriction b) Nonlinear conductance map measured from this constriction at 4.5 K. No clear Coulomb diamonds are visible; instead, there are narrow dips in the conductance at specific back gate voltages that extend over a broad range of source-drain biases.
}
  \label{fig:suppS5}
\end{figure}
\newpage


\begin{mcitethebibliography}{23}
\bibitem[Bibnote1()]{Bibnote1}
 Because two-wire measurements do not allow us to eliminate the effects of
  contact resistance, we base our mobility estimates on the slope of
  conductance traces at relatively low densities 10 V from the Dirac point
  where the graphene sheet resistance is still dominant over the contact
  resistances.\relax
\mciteBstWouldAddEndPunctfalse
\mciteSetBstMidEndSepPunct{\mcitedefaultmidpunct}
{}{\mcitedefaultseppunct}\relax
\EndOfBibitem
\bibitem[Bibnote2()]{Bibnote2}
 We calculate dot size based on a parallel plate capacitor model where the back
  gate defines one plate of the capacitor and the graphene sheet and nearby
  metal gates define the other plate, justified by the presence of conducting
  graphene sheet everywhere except for the two narrow strips defining the
  constriction. The presence in each of the samples discussed here of graphene
  or metal gates near the constriction may result in some screening of the back
  gate voltage, resulting in an underestimation of dot areas in each case. In
  data sets where dI/dV is measured with respect to side gate voltage, we
  calculate an equivalent back gate voltage using a separately measured
  multiplier.\relax
\mciteBstWouldAddEndPunctfalse
\mciteSetBstMidEndSepPunct{\mcitedefaultmidpunct}
{}{\mcitedefaultseppunct}\relax
\EndOfBibitem
\bibitem[Bibnote3()]{Bibnote3}
 The model we use to generate the expected double-dot behavior does not take
  into account the change in Fermi level with gate voltage that takes place in
  our system.\relax
\mciteBstWouldAddEndPunctfalse
\mciteSetBstMidEndSepPunct{\mcitedefaultmidpunct}
{}{\mcitedefaultseppunct}\relax
\EndOfBibitem
\end{mcitethebibliography}

\ifx\mcitethebibliography\mciteundefinedmacro
\PackageError{achemso.bst}{mciteplus.sty has not been loaded}
{This bibstyle requires the use of the mciteplus package.}\fi

\end{document}